\documentclass[aps,prl,reprint,twocolumn,superscriptaddress,floatfix,nofootinbib,longbibliography]{revtex4-2}
\usepackage{graphicx,amsmath,amsfonts,amssymb,amsthm,xr}
\usepackage{epsfig,amsmath,amssymb,color,dsfont,upgreek,physics}
\usepackage{mathrsfs}
\usepackage{mathtools}
\usepackage{bbold}
\usepackage{float}

\usepackage[bookmarks=true,colorlinks,linkcolor=OrangeRed,urlcolor=NavyBlue,citecolor=RoyalBlue]{hyperref}
\usepackage[dvipsnames]{xcolor}
\usepackage{orcidlink}

\usepackage{graphicx}
\usepackage{dcolumn}
\usepackage{bm}
\usepackage{color}

\begin{document}
\title{Generic Hilbert Space Fragmentation in Kogut--Susskind Lattice Gauge Theories}
\author{Anthony N.~Ciavarella$^{\orcidlink{0000-0003-3918-4110}}$}
\email{anciavarella@lbl.gov}
\affiliation{Physics Division, Lawrence Berkeley National Laboratory, Berkeley, California 94720, USA}
\author{Christian W.~Bauer$^{\orcidlink{0000-0001-9820-5810}}$}
\email{cwbauer@lbl.gov}
\affiliation{Physics Division, Lawrence Berkeley National Laboratory, Berkeley, California 94720, USA}
\affiliation{Department of Physics, University of California, Berkeley, Berkeley, CA 94720}

\author{Jad C.~Halimeh$^{\orcidlink{0000-0002-0659-7990}}$}
\email{jad.halimeh@physik.lmu.de}
\affiliation{Max Planck Institute of Quantum Optics, 85748 Garching, Germany}
\affiliation{Department of Physics and Arnold Sommerfeld Center for Theoretical Physics (ASC), Ludwig Maximilian University of Munich, 80333 Munich, Germany}
\affiliation{Munich Center for Quantum Science and Technology (MCQST), 80799 Munich, Germany}

\date{\today}

\begin{abstract}
    At the heart of quantum many-body physics lies the understanding of mechanisms that avoid quantum thermalization in an isolated system quenched far from equilibrium. 
    A prominent example is Hilbert space fragmentation, which has recently emerged as an ergodicity-breaking mechanism in constrained spin models. 
    Here, we show that Kogut--Susskind formulations of lattice gauge theories in $d+1$D ($d$ spatial and one temporal dimensions) give rise to Hilbert space fragmentation, and discuss possible implications for understanding continuum physics. 
    Our findings not only prove that lattice gauge theories are a natural platform for Hilbert space fragmentation, they also serve as a guide to the conditions under which these models can be faithfully used to infer the thermalization properties of quantum chromodynamics.
\end{abstract}

\maketitle

\textbf{\textit{Introduction.---}}An interacting disorder-free quantum many-body system quenched far from equilibrium is expected to \textit{quantum thermalize} at sufficiently long times, underpinning the Eigenstate Thermalization Hypothesis (ETH) \cite{Deutsch1991,Srednicki1994,Kaufman2016}. 
Local observables will eventually equilibrate in their dynamics to a value that can be predicted by a thermal ensemble \cite{Rigol_2008}. In the thermodynamic limit, this thermal ensemble is equivalent to an eigenstate of the quench Hamiltonian that has the same energy as the quench \cite{Rigol_review,Deutsch_review}. 
Such ergodic dynamics is fascinating given that it occurs in an isolated quantum system not connected to an external bath. Instead, the subsystems within act as baths relative to one another \cite{Eisert2015}.

Of great interest in the field of quantum many-body dynamics is the pursuit of mechanisms that break ergodicity, whereby the system does not thermalize for very long timescales. 
One of the earliest such mechanisms is many-body localization (MBL), where sufficiently strong disorder can prevent thermalization for long timescales and bring about \textit{localization} \cite{Gornyi2005,Basko2006,Nandkishore_review,Abanin_review,Alet_review}. 
Another mechanism is quantum many-body scarring (QMBS), which involves a polynomial number of special \textit{scar} eigenstates displaying area-law bipartite entanglement entropy and usually equally spaced in energy \cite{Bernien2017,Turner2018,Serbyn2020,MoudgalyaReview}. A recent prominent mechanism is Hilbert space fragmentation (HSF), where the Hilbert space fragments or \textit{shatters} into dynamically disconnected Krylov subspaces \cite{Sala2020ergodicity,Khemani2020localization}. 
This fragmentation is independent of underlying symmetries, but rather occurs even after resolving all exact global and local symmetries of the Hamiltonian. It is shown that the fragmentation is driven by emergent symmetries that arise in an effective Hamiltonian. The number of these Krylov sectors grows exponentially with system size \cite{Sala2020ergodicity,Khemani2020localization,Moudgalya2019,Moudgalya2022hilbert}.

Lattice gauge theories (LGTs) are a powerful framework conceived to enable non-perturbative calculations of quantum chromodynamics (QCD) and give insights into the nature of quark confinement \cite{Wilson1974},  and have also been shown to be greatly useful in condensed matter \cite{Wegner1971,Kogut_review,wen2004quantum}. 
There has also been a concerted effort to realize them on various quantum-hardware platforms \cite{Martinez2016,Klco2018,Goerg2019,Schweizer2019,Mil2020,Yang2020,Wang2021,Zhou2022,Wang2023,Zhang2023,Ciavarella2024quantum,cochran2024visualizingdynamicschargesstrings,gonzalezcuadra2024observationstringbreaking2,crippa2024analysisconfinementstring2,Ciavarella:2024lsp,de2024observationstringbreakingdynamicsquantum,liu2024stringbreakingmechanismlattice,Farrell:2023fgd,Farrell:2024fit,zhu2024probingfalsevacuumdecay,Ciavarella:2021nmj,Ciavarella:2023mfc,Ciavarella:2021lel,Gustafson:2023kvd,Gustafson:2024kym,Lamm:2024jnl,Farrell:2022wyt,Farrell:2022vyh,Li:2024lrl,Zemlevskiy:2024vxt,Lewis:2019wfx,Atas:2021ext,ARahman:2022tkr,Atas:2022dqm,Mendicelli:2022ntz,Kavaki:2024ijd,Than:2024zaj}, both for observing condensed-matter phenomena but also with the goal of eventually creating a complementary venue for probing high-energy physics \cite{Dalmonte_review,Pasquans_review,Zohar_review,Alexeev_review,aidelsburger2021cold,Zohar_NewReview,klco2021standard,Bauer_review,dimeglio2023quantum,Cheng_review,Halimeh_review,Cohen:2021imf,Lee:2024jnt,Turro:2024pxu}. 

In recent years, LGTs \cite{Wilson1974,Rothe_book} have emerged as a powerful platform for studying ergodicity-breaking phenomena. They have been shown to host a \textit{disorder-free} variant of MBL \cite{Smith2016,Brenes2018,gyawali2024observationdisorderfreelocalizationefficient}, arising upon quenching an initial state in a superposition of an extensive number of gauge superselection sectors. 
LGTs also display rich QMBS regimes \cite{Su2022,Desaules2022weak,Desaules2022prominent,aramthottil2022scar,desaules2024massassistedlocaldeconfinementconfined,Yao:2023pht,Ebner:2023ixq,Yao:2023gnm,Ebner:2024mee,Ebner:2024qtu} that seem to require the presence of the underlying gauge symmetry \cite{Halimeh2022robust}. 
When it comes to HSF, studies have focused on specific LGTs such as the $1+1$D $\mathrm{U}(1)$ quantum link model \cite{Chandrasekharan1997,Wiese_review} with a topological $\theta$-term \cite{Halimeh2022tuning,Cheng2022tunable}, where, in the limit of large strength of the latter, HSF can emerge \cite{Desaules2024ergodicitybreaking}. 
Another LGT displaying HSF is the lattice Schwinger model in the strong coupling limit \cite{jeyaretnam2025hilbertspacefragmentationorigin}. 
However, an overarching treatment of HSF in LGTs, particularly in determining under what conditions HSF can arise, has so far been lacking. Additionally, the dynamics of thermalization in gauge theories are relevant for understanding heavy ion collisions~\cite{Berges_review}. In these collisions, a quark-gluon plasma is produced and the process of thermalization plays a key role in the observed collective behavior~\cite{Mrowczynski:2016etf,Busza:2018rrf,Schlichting:2019abc}. There is also experimental evidence of thermalization occuring in hadronization during proton-proton and $e^+ e^-$ collisions~\cite{Hagedorn:1971mc,Becattini:2008tx,Becattini:2009sc,Kalaydzhyan:2015xba,Baker:2017wtt}. LGTs can provide non-perturbative input into these processes~\cite{Yao:2023pht,Ebner:2023ixq,Yao:2023gnm,Ebner:2024mee,Ebner:2024qtu,Hebenstreit2013,Buyens2016,Farrell:2024mgu,Florio:2023dke,Florio:2024aix,Grieninger:2024axp}, but a detailed understanding of thermalization on the lattice is necessary to apply this technique to the continuum limit of QCD.

In this Letter, we demonstrate that HSF occurs generically in LGTs. An analytic derivation based on the strong coupling expansion shows that LGTs with continuous compact gauge groups have local approximately conserved quantities even after gauge fixing. This results in a number of Krylov sectors that grow exponentially with system size. Numerical demonstrations are performed for U(1) and SU(2) LGTs. The potential implications for quantum simulations of LGTs and understanding thermalization in QCD are discussed.

\textbf{\textit{The Kogut--Susskind Hamiltonian.---}}The Kogut--Susskind Hamiltonian for a LGT with fermionic matter and compact gauge group $G$ is given by
\begin{align}
    \hat{H} & = \hat{H}_E + \hat{H}_B + \hat{H}_K +\hat{H}_M, \nonumber \\
    \hat{H}_E & = \frac{g^2}{2} \sum_{l} \hat{E}^2_{l},\,\,\,\hat{H}_M = \sum_{v} m\, \epsilon(v)  \, \hat{\chi}^\dagger_v \hat{\chi}_v, \nonumber \\
    \hat{H}_B & = - \frac{1}{2g^2} \sum_{p} \left(\hat{\Box}_p + \hat{\Box}^\dagger_p\right),  \nonumber \\
   \hat{H}_K &= \sum_{v,v'} \hat{\chi}^\dagger_v \eta_i(v) \Delta_i(v,v')\hat{U}(v,v') \hat{\chi}_{v'},
\end{align}
where $l$, $v$, and $p$ denote links, vertices, and plaquettes on the lattice, $\Delta_i(v,v') = (\delta_{v',v+i} - \delta_{v',v-i}) / 2$, $\epsilon(v)=(-1)^{\sum_i v_i}$, and $\eta_i(v) = (-1)^{\sum_{j<i}v_j}$. The electric energy operator on link $l$ is denoted by 
$\hat{E}^2_l$, the product of link operators around plaquette $p$ by $\hat{\Box}_p$, and the fermionic fields at vertex $v$ by $\hat{\chi}_v$. $g$ is the gauge coupling and $m$ is the fermion mass. $\hat{H}_E$ is the electric field energy, $\hat{H}_B$ is the plaquette term (corresponding to the magnetic energy in the continuum), $\hat{H}_M$ is the mass term for the fermion, and $\hat{H}_K$ is the kinetic term for the fermions that contains the minimal coupling to the gauge field.
Note that we have suppressed the dependence on color and the possibility of having fermions of different flavors. 

Even though we focus on the Kogut Susskind Hamiltonian displayed above, our results are in fact more general, and will allow flavor and position dependent fermion masses, position and flavor dependent kinetic fermion terms, as well as magnetic contributions involving larger closed loops. 
This will allow the formalism developed in this section to apply to different formulations of lattice fermions and the possible presence of a non-zero $\theta$ term. 
 
While several choices are possible for the basis of the Hilbert space~\cite{Hackett_2019,Alexandru_2019,Ji_2020,Bauer:2021gek,DAndrea:2023qnr,Grabowska:2024emw,Burbano:2024uvn,Jakobs:2023lpp,Muller:2023nnk,Ciavarella2024quantum,Fontana:2024rux,Gupta:2024gnw}, the most common is the so-called electric basis, for which the infinite-dimensional, gauge-invariant Hilbert space of this theory is spanned by electric basis vectors of the form $\ket{\mathcal{R}, \mathcal{G}, \mathcal{F}}$: ${\cal H} = {\rm span} \left( \ket{\mathcal{R}, \mathcal{G}, \mathcal{F}}\right)$, where $\mathcal{R}$ specifies all irreducible representations of the gauge group on each link, $\mathcal{G}$ specifies how the representations on neighboring links and fermionic matter add together to form a singlet representation on each site, and $\mathcal{F}$ specifies the fermion content on each site. 
More details are provided in the supplemental material.

In this basis, each contribution to the Hamiltonian can be written as
\begin{align}
    \hat H_X = &\!\!\!\!\sum_{\substack{{\mathcal{R},\mathcal{G},\mathcal{F}}\\{\mathcal{R}',\mathcal{G}',\mathcal{F}'}}} \!\!\!h_X(\mathcal{R}',\mathcal{G}',\mathcal{F}',\mathcal{R},\mathcal{G},\mathcal{F}) \ket{\mathcal{R}', \mathcal{G}', \mathcal{F}'} \bra{\mathcal{R}, \mathcal{G}, \mathcal{F}},
\end{align}
where the functions $h_X\equiv h_X(\mathcal{R}',\mathcal{G}',\mathcal{F}',\mathcal{R},\mathcal{G},\mathcal{F})=\bra{\mathcal{R}', \mathcal{G}', \mathcal{F}'}\hat H_X\ket{\mathcal{R}, \mathcal{G}, \mathcal{F}}$ with $X \in \{E,B,K,M\}$. 
One can therefore write
\begin{align}
    h_E & = \frac{g^2}{2} \sum_{l} C(\mathcal{R}_l) \delta_{\mathcal{R},\mathcal{R}'}\delta_{\mathcal{F},\mathcal{F}'}\delta_{\mathcal{G},\mathcal{G}'},\nonumber \\
    h_B & = - \frac{1}{2g^2} \sum_{p} \ p_p(\mathcal{R}',\mathcal{R},\mathcal{G}',\mathcal{G},\mathcal{F})\delta_{\mathcal{F},\mathcal{F}'},\nonumber \\
   h_K &= \sum_{v,v'} \eta_i(v) \, K(v,v',\mathcal{R}',\mathcal{R},\mathcal{G}',\mathcal{G},\mathcal{F}',\mathcal{F}),\nonumber \\
    h_M &=  m \, n_f(\mathcal{F}) \, \delta_{\mathcal{R},\mathcal{R}'}\delta_{\mathcal{F},\mathcal{F}'}\delta_{\mathcal{G},\mathcal{G}'} \,,
\end{align}
where $C(\mathcal{R}_l)$ is the Casimir of representation $\mathcal{R}_l$, and $n_f(\mathcal{F})$ counts the number of fermions present in configuration $\mathcal{F}$, taking into account the staggering factor $\epsilon(v)$.

\textbf{\textit{Fragmentation in KS Hamiltonians.---}}Note that in this basis, the electric and mass terms are diagonal, while the kinetic and plaquette terms are not. 
In particular, these two terms couple states with different representations $\mathcal{R}$ and $\mathcal{R}'$. 
The representations of a group of rank $N$ can be specified by a set of Dynkin indices $(m_1,m_2,\cdots,m_N)$, and the Casimir of a representation is quadratic in these indices. 
The change in representations mediated by the kinetic and plaquette terms is such that it amounts to a change in electric energy 
\begin{equation}
    \Delta E = \sum_{R'_l \in \mathcal{R}',R_l \in \mathcal{R}} \frac{g^2}{2} \big[C(R'_l) - C(R_l) \big] \ \ \ .
\end{equation}
The kinetic and plaquette terms will change Dynkin indices by $1$, so this change in energy will scale linearly with the Dynkin indices of the representations in $\mathcal{R}$ affected by the transition. 
On the other hand the matrix elements mediating the transition,  $p_p(\mathcal{R}',\mathcal{R},\mathcal{G}',\mathcal{G},\mathcal{F})$ and $K(v, v',\mathcal{R}',\mathcal{R},\mathcal{G}',\mathcal{G},\mathcal{F}',\mathcal{F})$, are bounded because the gauge group is compact. 
As a result, there will be transitions for which the change in the electric energy is large relative to the matrix elements of the plaquette and kinetic terms mediating the transition. 

One can perform an expansion in this hierarchy of terms using the Schrieffer-Wolff transformation~\cite{Foldy:1949wa,Schrieffer:1966abm,Bravyi2011}.
The details are presented in the supplemental material, and it results in an effective Hamiltonian
valid below some large scale $\mathcal{E}$. 
To write the effective Hamiltonian, we separate the representations on the links of the lattice into those with Casimir less than some value $\mathcal{E}^2$ and those with Casimir above $\mathcal{E}^2$. 
The Hilbert space can then be written as ${\cal H}_\mathcal{E} = {\rm span} \big( \ket{\mathcal{R}_{<\mathcal{E}}, \mathcal{R}_{>\mathcal{E}}, \mathcal{G}, \mathcal{F}}\big)$, where $\mathcal{R}_{<\mathcal{E}}$ is the set of all representations on the links with Casimir below $\mathcal{E}^2$:  $\mathcal{R}_{<\mathcal{E}}= \{R_l : \forall R_l \in \mathcal{R}, C(R_l) < \mathcal{E}^2 \}$, while $\mathcal{R}_{>\mathcal{E}}$ is the set with $C(R_l) \geq \mathcal{E}^2$.
The matrix elements of the effective Hamiltonian are then given by $h_X^\mathcal{E} = h_X \, \delta_{\mathcal{R}'_{>\mathcal{E}},\mathcal{R}_{>\mathcal{E}}}$. In other words, the effective Hamiltonian is given by the original Hamiltonian, but with all transitions between representations with Casimirs above $\mathcal{E}^2$ removed. This effective Hamiltonian will have errors that are order $1/(g^6\mathcal{E})$ from the truncation of the plaquette term and errors that are order $1/(g^2\mathcal{E})$ from the truncation of the kinetic term. 

This implies that the Hilbert space separates into a set of subspaces, each of which has a fixed set of representations in $\mathcal{R}_{\mathcal{E}}$. 
The number of such subspaces is exponential in the number of links in the lattice, and therefore exponential in the volume of the lattice. 
The dimensionality of most of these subspaces is still exponential in the number of links in the set $\mathcal{R}_{<\mathcal{E}}$, and therefore exponential in the volume of the lattice.
The effective Hamiltonian is block-diagonal with respect to these subspaces, demonstrating emergent symmetries. 
This leads to Hilbert space fragmentation in the same way as in the Ising and Hubbard models as found in~\cite{Yoshinaga:2021oqd,Kwan:2023kjp}.

\textbf{\textit{U(1) lattice gauge theory.---}}
To illustrate the ideas that go into this argument, we will look at a $\mathrm{U}(1)$ LGT on a ladder without matter. 
Using Gauss's law and constraining ourselves to states connected to the electric vacuum, electric basis states can be specified by the electric field present on the upper links of the ladder. The Kogut--Susskind Hamiltonian is given by
\begin{align}
    \hat{H} &= \sum_l \frac{g^2}{2} \hat{E}_l^2 - \frac{1}{2g^2} \sum_p \left(\hat{U}_p + \hat{U}^\dagger_p \right) \,.
\end{align}
The representations of $\mathrm{U}(1)$ are labeled by integers, and the Hilbert space is therefore spanned by a set of integers for each link, constrained by Gauss' law such that the integers of all links emanating from a given vertex add to zero.  
In this basis the operators are given by
\begin{align}
    \hat{E}_l &= \sum_{n_l\in \mathbb{Z}} n_l \ket{n_l}\bra{n_l} \nonumber \\
    \hat{U}_p &= \prod_{l\in p} \left[\sum_{n_l\in \mathbb{Z}}  \ket{n_l+1}\bra{n_l}\right] \,,
\end{align}
where the product in $U_p$ is over the 4 links of the plaquette. 
The basis states of a lattice with length $L$ is then specified by $3L+1$ integers, with Gauss' law providing $2L$ constraints, for a total of $L$ independent sets of integers. 

Following the arguments of the previous section, for sufficiently large $\mathcal{E}$, the system is described by the effective Hamiltonian, which only includes the operators $\hat U_p$ for those plaquettes that have $n_l < \mathcal{E}$ for all 4 links on the plaquette. 
As before, this gives a number of conserved quantities that is proportional to the size of the lattice, with the dimensionality of the Hilbert space corresponding to each sector being exponential in lattice size. 
This is sufficient to guarantee that Hilbert space fragmentation occurs. 

As a numerical check on these arguments, an $L=4$ system was simulated with the electric fields truncated at $|n|<5$. 
In this simulation, we choose to work with an explicitly gauge-fixed Hilbert space, with all states satisfying Gauss's law. 
By restricting to states dynamically connected to the state with all electric fields zero, the horizontal links on a plaquette are constrained to have the same magnitude of electric fields with opposite sign. The electric fields on the vertical links are fixed by Gauss's law. This allows electric basis states to be specified by the electric field on the upper link of each plaquette, denoted $\hat{E}_p$. Explicitly, the basis states are $\ket{n_1,n_2,\cdots,n_L}$ where each $n_p$ is an integer. 
The Hamiltonian in this basis is given by
\begin{align}
    \hat{H} &= \sum_p \bigg\{g^2 \left[ \hat{E}_p^2 + \frac{1}{2}\Big(\hat{E}_p - \hat{E}_{p+1}\Big)^2\right]- \frac{1}{2g^2} \left(\hat{U}_p + \hat{U}^\dagger_p \right)\bigg\},
\end{align}
where $\hat{E}_p = \sum_{n_p\in \mathbb{Z}} n_p \ket{n_p}\bra{n_p}$ and $\hat{U}_p = \sum_{n\in \mathbb{Z}}  \ket{n_p+1}\bra{n_p}$.
The term $\hat E_p^2$ denotes the sum of the electric energy of the two horizontal links of each plaquette (which are identical in this ladder configuration), while the term $(\hat{E}_p - \hat{E}_{p+1})^2 / 2$ is the electric energy in the (right) vertical link of each plaquette.

\begin{figure}
    \centering
    \includegraphics[width=8.6cm]{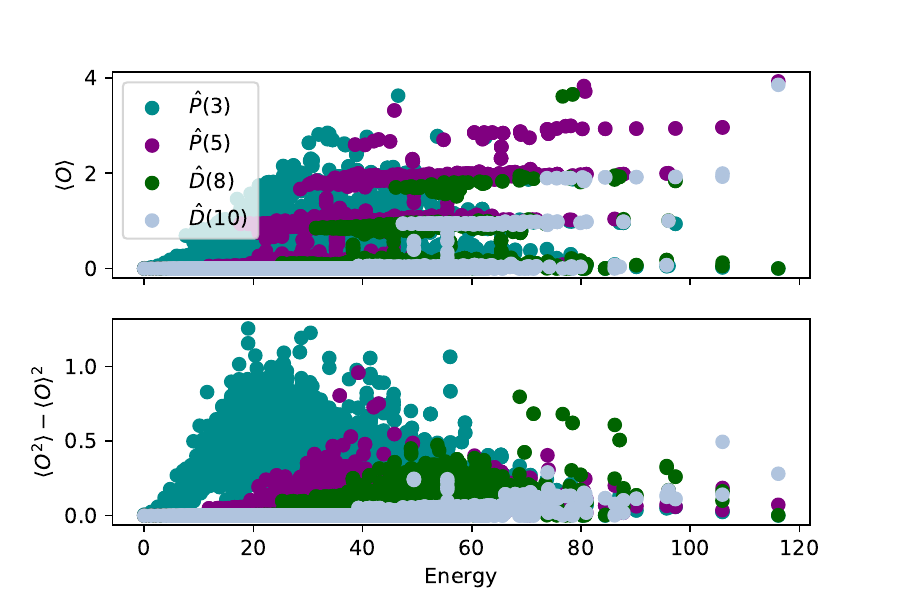}
    \caption{Expectation and variance of $\hat{P}(s)$ and $\hat{D}(s)$ for eigenstates in the momentum zero sector of a $L=4$ $\mathrm{U}(1)$ LGT.
    The upper plot shows the expectation of the operator and the lower plot shows the variance.}
    \label{fig:LocalProjector}
\end{figure}

The expectation of the operators
\begin{align}
    \hat{P}(s) & = \sum_{p} \ket{s_p} \bra{s_p}, \nonumber \\
    \hat{D}(s) & = \sum_{p} \sum_{n,k \ \text{s.t.} \abs{n-k}=s} \ket{n_p} \bra{n_p} \otimes \ket{k_{p+1}} \bra{k_{p+1}},
\end{align}
for energy eigenstates in the zero momentum sector of the theory was computed. 
These operators $\hat P(s)$ ($\hat D(S)$) measure the number of horizontal (vertical) links with given representation $s$.
If $s$ is large enough to specify different fragmentation sectors, then eigenstates of $\hat{H}$ should also be eigenstates of $\hat{P}(s)$ and $\hat{D}(s)$, such that the expectation values of the operators should be given by integer values, with vanishing variance.
Figure~\ref{fig:LocalProjector} shows the expectation and variance of $\hat{P}(s)$ and $\hat{D}(s)$ for zero momentum eigenstates with $g=0.6$. 
At this coupling, it can be seen that as $s$ increases, the expectation of these operators concentrates around integer values and the variance decreases. 
This indicates that as $s$ is increased, they become closer to a conserved quantity, as expected from the argument for fragmentation.

\begin{figure}
    \centering
    \includegraphics[width=8.6cm]{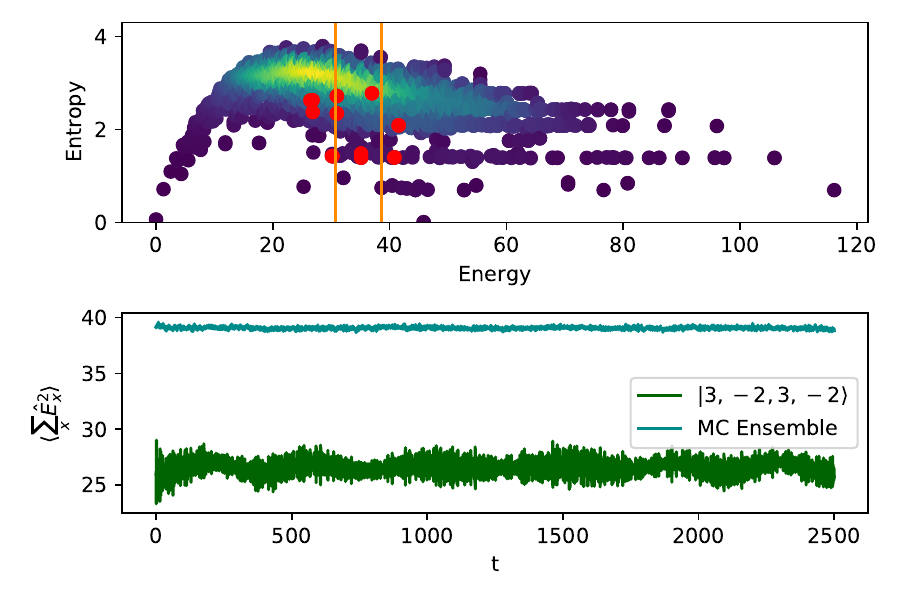}
    \caption{The upper panel shows the half-chain entropy of the zero momentum energy eigenstates of a $\mathrm{U}(1)$ $L=4$ plaquette chain with $g=0.6$. 
    The red points show energy eigenstates that have $\abs{\bra{E_n}\ket{3,-2,3,-2}}^2>0.05$ and the states between the orange lines are in the microcanonical ensemble state $\ket{\psi_{MC}}$. The lower panel shows the expectation of $\sum_x\hat{E}^2_x$ as a function of time $t$ for the initial states $\ket{3,-2,3,-2}$ and $\ket{\psi_{MC}}$.}
    \label{fig:U1Freeze}
\end{figure}

The half-chain entropy of the zero momentum eigenstates at this coupling is shown in Fig.~\ref{fig:U1Freeze}. In this figure, there are lines of states with the same entropy with different energies. 
These are eigenstates that are translationally invariant superpositions of frozen product states. The non-zero entanglement entropy comes from these product states being projected into the zero momentum sector.
To demonstrate the freezing of dynamics, the state $\ket{3,-2,3,-2}$ was evolved in time. Figure~\ref{fig:U1Freeze} shows the evolution of $\ket{\psi_0}=\ket{3,-2,3,-2}$ and a microcanonical state, $\ket{\psi_{MC}}$, at the same energy. 
$\ket{\psi_{MC}}$ is given by an equal superposition of energy eigenstates in the interval $[E-\delta E,E+\delta E]$ where $E=\bra{\psi_0}\hat{H}\ket{\psi_0}$ and $\delta E=\sqrt{\bra{\psi_0}\hat{H}^2\ket{\psi_0}-E^2}$. 
This figure shows that the expectation of $\sum_x \hat{E}_x^2$ with initial state $\ket{\psi_0}$ does not thermalize to the microcanonical expectation within the simulated timescales, as would be expected from HSF.

\begin{figure}
     \centering
     \includegraphics[width=8.6cm]{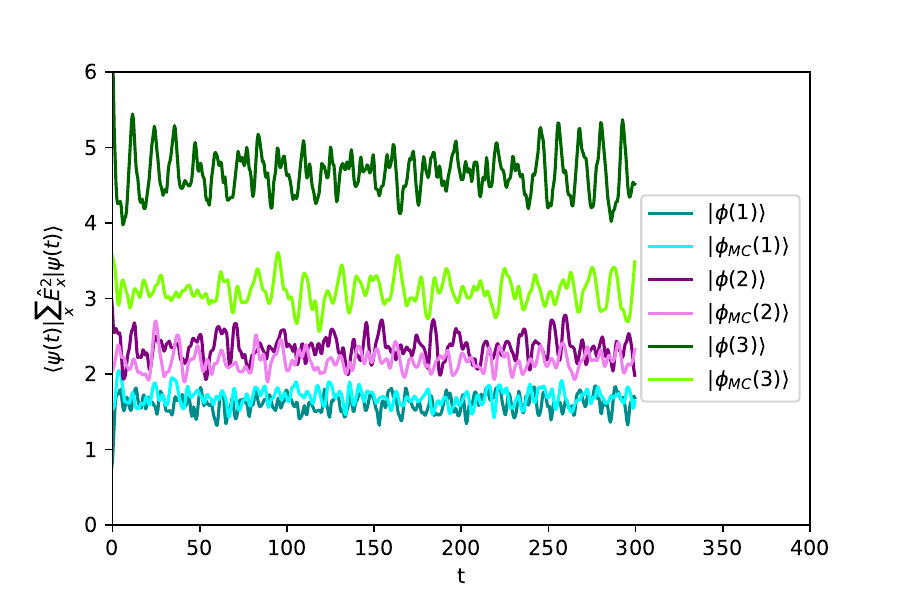}
     \caption{Evolution of the electric energy as a function of time on a $6$ site lattice with open boundary conditions. 
     The dark lines show the evolution of states with different numbers of quark pairs placed on top of the electric vacuum, and the light lines show the evolution of the corresponding microcanonical ensemble states.}
     \label{fig:SU2Fragmentation}
 \end{figure}

\textbf{\textit{SU(2) lattice gauge theory.---}}
An $\mathrm{SU}(2)$ LGT in $1+1$D will be used to demonstrate Hilbert space fragmentation in a theory with matter. 
The Hamiltonian for a theory with a single flavor of staggered fermions is given by
\begin{align}
    \hat{H} &= \hat{H}_E + \hat{H}_K + \hat{H}_M, \nonumber \\
    \hat{H}_E &= \sum_{v,a} \frac{g^2}{2} \left(\hat{E}^a_{v}\right)^2,\,\,\,\hat{H}_M = m \sum_{v,a} \epsilon(v) \hat{\chi}^\dagger_{v,a} \hat{\chi}_{v,a},  \nonumber \\
    \hat{H}_K &= \sum_{v,a,b} \frac{1}{2}\hat{\chi}_{v,a}^\dagger \Delta(v,v') \hat{U}_{a,b}(v,v')  \hat{\chi}_{v',b},
\end{align}
where $\chi_{v,a}$ is a fermionic field at site $v$ with color $a$, $\hat{U}_{a,b}(v,v')$ is an SU(2) parallel transporter between sites $v$ and $v'$, and $\hat{E}_{v}^a$ is the corresponding electric field operator. 
Using Gauss's law and open boundary conditions, the gauge fields can be integrated out, yielding
\begin{align}
    \hat{H}_E &= \sum_{v,a} \frac{g^2}{2} \left( \sum_{v'<v,c,c'} \frac{1}{2} \hat{\chi}_{v',c'}^\dagger \sigma^a_{c',c}  \hat{\chi}_{v',c}\right)^2, \nonumber \\
    \hat{H}_K &= \sum_{v,a} \frac{1}{2}\hat{\chi}_{v+1,a}^\dagger  \hat{\chi}_{v,a} + \text{h.c.},
\end{align}
with $\hat H_M$ staying unchanged. 
Here $\sigma_a$ is a Pauli matrix. In this theory, regions with large electric fields will be expected to freeze, and local observables should fail to thermalize to the microcanonical expectation. Note that in this theory, matter must be present in gauge-invariant states to provide a source for the gauge fields. HSF will be demonstrated using states of the form
\begin{align}
    & \ket{\phi(v,D)}  = \nonumber \prod_{s=0}^{D-1} \\
    & \left(\sum_a \frac{1}{\sqrt{2}}\hat{\chi}^\dagger_{x-s,a} \hat{\chi}_{x+s+1,a} + \frac{1}{\sqrt{2}}\hat{\chi}_{x-s,a} \hat{\chi}^\dagger_{x+s+1,a}\right) \ket{0} \,,
\end{align}
where $\ket{0}$ is the electric vacuum with all anti-fermion sites filled and $D$ is the number of fermion anti-fermion pairs.
These states will have electric fields with representations varying from $j=0$ to $j=\frac{D}{2}$ on the link between sites $v$ and $v+1$. 
If $D$ is large enough, this state will have support on multiple sectors of the theory and fail to thermalize to the microcanonical expectation. 
For smaller $D$, the state should thermalize to the microcanonical ensemble. This will be demonstrated numerically on a lattice with $6$ staggered sites with open boundary conditions, $g=1$ and $m=0.1$. Figure~\ref{fig:SU2Fragmentation} shows the expectation of the electric energy as a function of time for the initial states $\ket{\phi(2,D)}$ and the corresponding microcanonical ensemble states. 
For $D=1$ and $D=2$, the expectation of the electric energy is consistent with the expectation from the microcanonical ensemble. For $D=3$, the system does not thermalize to the microcanonical expectation, consistent with the arguments for Hilbert space fragmentation.

\textbf{\textit{Discussion.---}} 
Hilbert space fragmentation in LGTs prevents the dynamics of the gauge theory from changing the states of links that are in electric basis states with large Casimirs. 
Importantly, if one initializes a system with all links in low-lying representations, the system at later times will not populate links with representations above some cutoff value.  
This is relevant for performing quantum simulations of LGTs, as most approaches work with a truncation of the Hilbert space in an electric basis. HSF implies that for a simulation performed at fixed $g$, one only needs to raise the truncation of electric basis states to the point where HSF occurs to have agreement with the untruncated Hamiltonian. 
As $g$ sets the lattice spacing, this can be understood as the lattice spacing setting a cutoff on the electric energy density.
An interesting question is the behavior at extremely long times; it is possible for there to be tunneling between the fragmentation sectors, as the fragmentation is exact only in an effective Hamiltonian. At the timescales accessed in our simulations, this was not observed. These simulations were performed on moderate system sizes to enable numerical studies of the full spectrum of the theory. As the system size is increased, it is expected that these signatures of HSF will remain, as the effective Hamiltonian description is not limited by system size.

The presence of HSF also impacts the difficulty of classically simulating LGT dynamics. 
Naively, as the energy density of a system is increased, one would expect the computational costs of simulations to increase. However, due to HSF, the dynamics of LGTs will freeze above some cutoff in energy density and become easy to simulate. 
The computational costs of performing a simulation only grow as $g$ is lowered, i.e., as the continuum limit is approached. This could be used to perform a quantum simulation with verifiable advantage by performing the verification with the dynamics of a state with large electric fields, similar to the proposal in~\cite{Hartse:2024qrv}.

One might be worried about the implications of HSF for predictions in the continuum theory.
States with large electric flux flowing around a plaquette would correspond to a glueball, whose stability would be guaranteed by the freezing of the electric flux. 
Additionally, due to HSF freezing the position of the electric flux, these glueballs would be unable to move.
However, the arguments for HSF are based on the validity of the strong coupling expansion for states with links with large electric energy. 
The strong coupling expansion fails as the continuum limit is approached, which suggests that the HSF found in this work is a lattice artifact. 
The fact that the lattice theory only reproduces the continuum theory below a UV cutoff is another indication that the freezing out of high-energy states is a lattice artifact. 
The results of this paper show that one needs to be careful when using LGTs to make predictions about thermalization properties of QCD.

\medskip

\begin{acknowledgements}
\footnotesize{\textbf{\textit{Acknowledgments.---}}A.N.C.~and C.W.B.~were supported by the U.S. Department of Energy, Office of Science, National Quantum Information 
Science Research Centers, Quantum Systems Accelerator and the U.S.~Department of Energy, Office of Science under contract DE-AC02-05CH11231, primarily through Quantum Information Science Enabled Discovery (QuantISED) for High Energy Physics (KA2401032). 
J.C.H.~acknowledges funding by the Max Planck Society, the Deutsche Forschungsgemeinschaft (DFG, German Research Foundation) under Germany’s Excellence Strategy – EXC-2111 – 390814868, and the European Research Council (ERC) under the European Union’s Horizon Europe research and innovation program (Grant Agreement No.~101165667)—ERC Starting Grant QuSiGauge. This work is part of the Quantum Computing for High-Energy Physics (QC4HEP) working group.}
\end{acknowledgements}

\bibliographystyle{apsrev4-1}
\bibliography{biblio}

\pagebreak

\onecolumngrid

\setcounter{equation}{0}
\setcounter{figure}{0}
\setcounter{table}{0}

\renewcommand{\theequation}{S\arabic{equation}}
\renewcommand{\thefigure}{S\arabic{figure}}

\begin{center}
\textbf{\large Generic Hilbert Space Fragmentation in Kogut--Susskind Lattice Gauge Theories: Supplemental Material}
\end{center}

\section{Fragmentation in the Kogut-Susskind Hamiltonian}
\subsection{Basis Setup (Pure Gauge)}
In this section, it is shown how to construct a gauge invariant electric basis for a lattice gauge theory without matter. Electric basis states of the Kogut-Susskind Hamiltonian on a $d$ dimensional lattice with a compact gauge group $G$ are given by states of the form $\ket{R,r_L,r_R}$ on each link, where $R$ is an irreducible representation of $G$, $r_L$ is a component of $R$ associated with the left side of the link and $r_R$ is a component of $R$ associated with the right side of the link. 
Each link has a parallel transporter, and left and right electric field operators defined by
\begin{align}
    \hat{U}^{F, \mu}_{\alpha\beta}(\Vec{x})\ket{R,r_L,r_R} &= \sum_{R',r_L',r_R'}\sqrt{\frac{\text{Dim}(R)}{\text{Dim}(R')}} C^{R' r_L'}_{F \beta;R r_L} C^{R' r_R'}_{F \alpha;R r_R} \ket{R',r_L',r_R'} \nonumber \\
    \hat{E}^{a, \mu}_L\ket{R,r_L,r_R}(\Vec{x}) &= \sum_{r_L'} -T^a_{r_L',r_L} \ket{R,r_L',r_R} \nonumber \\
    \hat{E}^{a, \mu}_R\ket{R,r_L,r_R}(\Vec{x}) &=  \sum_{r_R'} T^a_{r_R',r_R} \ket{R,r_L,r_R'} \ \ \ ,
\end{align}
where $F$ is an irreducible representation of $G$, $\Vec{x}$ specifies a lattice site, $\mu$ specifies a direction on the lattice, $T^a_{ij}$ are the generators of $G$, and $C^{A,a}_{B,b;C,c}$ are Clebsh-Gordan coefficients of $G$ describing how irreps $A$ and $B$ combine to form $C$. 
Gauge invariant operators can be formed by squaring the electric field operator, i.e.
\begin{equation}
    \hat{E}^2_{\mu}(\Vec{x}) = \sum_a \hat{E}^{a, \mu}_L \hat{E}^{a, \mu}_L = \sum_a \hat{E}^{a, \mu}_R \hat{E}^{a, \mu}_R \ \ \ ,
\end{equation}
or by taking the product of parallel transporters around a closed curve $C$ on the lattice, i.e.
\begin{equation}
    \hat{\Box}_C = \sum_{s} \prod_{\Vec{x}_n,\mu_n \in C} \hat{U}^{F, \  \mu_n}_{s_{n+1},s_n}(\Vec{x}_n) \ \ \ .
\end{equation}
Typically $F$ is taken to be the fundamental representation of the gauge group, but in principle, any irreducible representation can be used. 
Given a set of closed curves $\mathcal{C}$, the Kogut-Susskind Hamiltonian is 
\begin{equation}
    \hat{H}_{KS} = \frac{g^2}{2} \sum_{\Vec{x},\mu} \hat{E}^2_{\mu}(\Vec{x}) - \frac{1}{2g^2} \sum_{C \in \mathcal{C}} \hat{\Box}_C + \hat{\Box}^\dagger_C 
    \label{eq:KogutS}
\end{equation}
where $g$ is the coupling constant. The Kogut-Susskind Hamiltonian is most often defined on a $d$ dimensional square lattice and $\mathcal{C}$ is usually taken to be the set of the smallest closed loops on the lattice (called plaquettes), however, it is possible to consider more general lattices and plaquette operators. 
The Kogut-Susskind Hamiltonian has a gauge symmetry generated by the sum over electric operators at each site. This gauge symmetry can be enforced by restricting the electric fields at each site to add up to an irreducible representation. 
Typically, gauge fields are chosen to add up to the singlet representation, but other representations can be chosen which corresponds to placing external charges on the lattice. 
This work will focus on the charge zero sector, but the results should also generalize to other charge sectors. 
In a basis where Gauss's law is enforced (i.e. electric fields at all sites add up to the singlet representation), gauge invariant basis states are given by $\ket{\mathcal{R}, \mathcal{G}}$ where $\mathcal{R}=\{R_l:l\in \text{Links} \}$, $\mathcal{G}=\{G_i:i\in \text{Sites} \}$, $R_l$ specifies the irreducible on link $l$ and $G_i$ specifies how the representations on the links connected to site $i$ add up to form a singlet state. 
In this gauge invariant basis, the electric energy operator for a link $l$ is 
\begin{equation}
    \hat{E}^2_l \ket{\mathcal{R},\mathcal{G}} = C(R_l)\ket{\mathcal{R},\mathcal{G}} \ \ \ ,
\end{equation}
where $C(R_l)$ is the Casimir of the representation on link $l$. 
The plaquette operator changes the link representations and site labels for each link and site along the curve defining the plaquette and the matrix elements are given by
\begin{equation}
   \bra{\mathcal{R'}, \mathcal{G'}} \hat{\Box}_C \ket{\mathcal{R}, \mathcal{G}} = p_C(\mathcal{R'}, \mathcal{G'},\mathcal{R}, \mathcal{G}) = \prod_{i \in C} f_i(S_i(\mathcal{R'}), S(\mathcal{R}),G_i)\ \ \ ,
\end{equation}
where $i$ is a site along the curve $C$, $S_i$ is the set of irreps on the links connected to site $i$, and $f_i$ is a vertex factor given by a contraction of Clebsch-Gordan coefficients that depends on the details of how $G_i$ is constructed. 
Note that the plaquette operator is defined by the trace of a product of group operators so the matrix elements of the plaquette operator are upper bounded by the dimension of the representation $F$ used to define the parallel transporters, i.e.
\begin{equation}
    \abs{\bra{\mathcal{R'}, \mathcal{G'}} \hat{\Box}_C \ket{\mathcal{R}, \mathcal{G}}} \leq \text{dim}(F) \ \ \ .
\end{equation}
With this basis, the Kogut-Susskind Hamiltonian for a pure gauge theory can be written as
\begin{align}
    \hat H_X = &\!\!\!\!\sum_{\substack{{\mathcal{R},\mathcal{G}}\\{\mathcal{R}',\mathcal{G}',}}} \!\!\!h_X(\mathcal{R}',\mathcal{G}',\mathcal{F}',\mathcal{R},\mathcal{V},\mathcal{F}) \ket{\mathcal{R}', \mathcal{G}', \mathcal{F}'} \bra{\mathcal{R}, \mathcal{G}, \mathcal{F}},
\end{align}
where the functions $h_X\equiv h_X(\mathcal{R}',\mathcal{G}',\mathcal{R},\mathcal{G})=\bra{\mathcal{R}', \mathcal{G}'}\hat H_X\ket{\mathcal{R}, \mathcal{G}}$ with $X \in \{E,B\}$ are given by
\begin{align}
h_E(\mathcal{R}',\mathcal{G}',\mathcal{R},\mathcal{G}) & = \bra{\mathcal{R}', \mathcal{G}'}\hat H_E\ket{\mathcal{R}, \mathcal{G}} = \frac{g^2}{2} \delta_{\mathcal{R}',\mathcal{R}} \delta_{\mathcal{G}',\mathcal{G}} \sum_{R_l \in \mathcal{R}} C(R_l) \nonumber \\
h_B(\mathcal{R}',\mathcal{G}',\mathcal{R},\mathcal{G}) & =  \bra{\mathcal{R}', \mathcal{G}'}\hat H_B\ket{\mathcal{R}, \mathcal{G}} = -\frac{1}{2g^2} \sum_{C\in\mathcal{C}} p_C(\mathcal{R'}, \mathcal{G'},\mathcal{R}, \mathcal{G}) \ \ \ .
\end{align}
 
\subsection{The Schrieffer--Wolf Expansion and Derivation of the Effective Hamiltonian}
An effective Hamiltonian that separates out states with large electric energy densities will be needed to demonstrate fragmentation. This will be derived using the Schrieffer-Wolff transformation, which will be reviewed in this section. For a generic Hamiltonian
\begin{equation}
    \hat{H} =\hat{H}_0 + \hat{V} \ \ \ ,
\end{equation}
Schrieffer-Wolff perturbation theory constructs a unitary transformation with a perturbative expansion in powers of $\hat{V}$ that diagonalizes $\hat{H}$. Applying this unitary transformation to the Hamiltonian and truncating at some power in $\hat{V}$ will give an effective Hamiltonian. In this expansion, $\hat{H}_0$ can be thought of as a diagonal term, and $\hat{V}$ can be thought of as the off-diagonal, so the generator of the unitary will be chosen to cancel odd powers of $\hat{V}$ in the effective Hamiltonian.

The Schrieffer-Wolff transformation is generated by an anti-Hermitian operator $\hat{S}$ and up to second order in $\hat{V}$, an effective Hamiltonian is given by
\begin{equation}
    \hat{H}_{eff} = e^{\hat{S}}\left(\hat{H}_0 + \hat{V}\right) e^{-\hat{S}} \approx \hat{H}_0 + [\hat{S},\hat{H_0}] + \frac{1}{2}[\hat{S},[\hat{S},\hat{H_0}]] + \hat{V} + [\hat{S},\hat{V}]+\mathcal{O}(\hat{V}^3)
\end{equation}
To remove the off-diagonal pieces of the Hamiltonian to leading order, we require
\begin{equation}
    [\hat{S},\hat{H_0}] + \hat{V} = 0 \ \ \ .
\end{equation}
Using this to fix $\hat{S}$, the effective Hamiltonian to second order is
\begin{equation}
    \hat{H}_{eff} = \hat{H}_0 +\frac{1}{2}[\hat{S},\hat{V}] \ \ \ .
\end{equation}
In the eigenbasis of $\hat{H}_0$, the second order term is
\begin{equation}
    \bra{i}\frac{1}{2}[\hat{S},\hat{V}]\ket{f} = \frac{1}{2}\sum_k V_{ik} V_{kf} \left(\frac{1}{E_i - E_k} - \frac{1}{E_k - E_f}\right) \ \ \ ,
\end{equation}
where $E_i$ are the eigen-energies of $\hat{H}_0$ and $V_{nk}$ are the matrix elements of $\hat{V}$ in the eigenbasis of $\hat{H}_0$. This is a valid expansion whenever the matrix elements of $\hat{V}$ are small relative to the change in energy of the $\hat{H}_0$ eigenstates coupled by $\hat{V}$.

\subsection{Application to Kogut-Susskind and Fragmentation}
The electric term of the Kogut-Susskind Hamiltonian is given by the sum over Casimirs of the representations on the links. Generically, the Casimir is quadratic in the Dynkin indices, and the matrix elements of the plaquette operators will be upper-bounded by the dimension of the representation used to define the parallel transporters. This means that for basis states with sufficiently high dimensional irreducible representations, the change in electric energy for states coupled by a plaquette operator will be large relative to the plaquette operator matrix elements. This allows for the system to be described by an effective Hamiltonian where links with high dimensional representations are frozen. Explicitly for some $\mathcal{E}$, there will be representations $R$ with a Casimir that satisfies $C(R) > \mathcal{E}^2$. When a plaquette operator changes $R$ to $R'$ the change in Casimir will be $\Omega(\mathcal{E})$. 
To write the effective Hamiltonian, we separate the representations on the links of the lattice into those with Casimir less than some value $\mathcal{E}^2$ and those with Casimir above $\mathcal{E}^2$. 
The basis states can be rewritten as $\ket{\mathcal{R}_{<\mathcal{E}}, \mathcal{R}_{>\mathcal{E}},\mathcal{G}}$, where $\mathcal{R}_{<\mathcal{E}}$ is the set of all representations on the links with Casimir below $\mathcal{E}^2$, while $\mathcal{R}_{>\mathcal{E}}$ is the set with $C(R_l) \geq \mathcal{E}^2$.
The Kogut-Susskind Hamiltonian can be split into a block diagonal piece, $\hat{H}_0$ and an off-diagonal piece $\hat{V}$ by separating out the matrix elements that couple states with large electric fields. The matrix elements of $\hat{H}_0$ are given by $h_X^\mathcal{E} = h_X \, \delta_{\mathcal{R}'_{>\mathcal{E}},\mathcal{R}_{>\mathcal{E}}}$ and $\hat{V} = \hat{H}_{KS} - \hat{H}_0$.
Using the Schrieffer-Wolff formalism described in the previous section, $\hat{H}_0$ is an effective Hamiltonian that describes the system with corrections that are order $\mathcal{O}\left(\frac{1}{g^6 \mathcal{E}}\right)$. These corrections come from the second order term in the Schrieffer-Wolff expansion. Note that because continuous gauge groups have an infinite number of representations with arbitrarily large Casimirs, there will always be some $\mathcal{E}$ where this is a small correction. Under $\hat{H}_0$, operators of the form $\ket{R_l}\bra{R_l}$ for representations with $C(R_l)\geq \mathcal{E}^2$ are conserved quantities. The number of conserved quantities of this form is proportional to the lattice size. This results in Hilbert space fragmentation where different symmetry sectors have frozen subregions of the lattice where links carry large electric energies. Note that in these frozen sectors, only a subregion of the lattice will be frozen out and the rest of the lattice will have non-trivial dynamics.

\subsection{Inclusion of Matter}
The basis introduced in the previous sections can be generalized to include fermionic matter. There are a number of different formulations of lattice fermions and in this work staggered fermions will be used. However, our results should apply to any type of lattice fermion and with multiple flavors of fermions. When matter is coupled to gauge fields, a kinetic and mass term for the matter are added to the Hamiltonian,
\begin{align}
     \hat{H}_K &= \sum_{v,v'} \hat{\chi}^\dagger_v \eta_i(v) \Delta_i(v,v')\hat{U}(v,v') \hat{\chi}_{v'} \nonumber \\
     \hat{H}_M &= \sum_{v} m\, \epsilon(v)  \, \hat{\chi}^\dagger_v \hat{\chi}_v
\end{align}
where $\hat{\chi}_v$ is a fermionic field at site $v$, $\hat{U}(v,v')$ is a product of parallel transporters connecting site $v'$ to site $v$, $\Delta_i(v,v') = (\delta_{v',v+i} - \delta_{v',v-i}) / 2$ where $i$ is a direction along the spatial lattice, $\eta_i(v) = (-1)^{\sum_{k=1}^{i-1} v_i}$ is a phase that comes from spin diagonalizing the hopping term, $m$ is the fermion mass, and $\epsilon(v) = (-1)^{\sum_{k=1}^{d} v_i}$. All color indices have been suppressed in this expression. The Hilbert space of a single site can be described by basis states
\begin{equation}
    \ket{c_1,c_2,\cdots, c_k} = \prod_j \hat{\chi}^\dagger_{v,c_j} \ket{0}\ \ \ ,
\end{equation}
where $\ket{0}$ is a state with no fermions present. The color indices of the state can be combined into an irreducible representation of the gauge group. Therefore, the basis states including matter are given by $\ket{\mathcal{R}, \mathcal{G}, \mathcal{F}}$ where $\mathcal{F} = \{(C_i,n_i),i\in\text{Sites}\}$, $C_i$ are representations of the gauge group, and $n_i$ is the number of fermions at site $i$. The degree of freedom describing which component of $C_i$ the fermions are in can be removed by enforcing Gauss's law in the same way that this degree of freedom was removed for the gauge fields. With this basis, the mass term is diagonal and the kinetic term is off-diagonal, coupling states with different $\mathcal{R}$, $\mathcal{G}$, and $\mathcal{F}$. The argument for Hilbert space fragmentation proceeds in the same manner as the case without matter. With the inclusion of matter, the effective Hamiltonian $\hat{H}_0$ will have additional corrections that are order $\mathcal{O}\left(\frac{1}{g^2 \mathcal{E}}\right)$.

\end{document}